\documentclass[aps,showpacs,twocolumn,superscriptaddress,nofootinbib,floatfix,section 10pt]{revtex4-1}
\usepackage{amsfonts,amssymb,stmaryrd,latexsym,amsmath,braket}
\usepackage{graphicx,subfigure}
\usepackage{comment}
\usepackage{times}
\usepackage{slashed}
\usepackage{bm}
\usepackage{braket}
\usepackage{appendix}
\usepackage{enumitem}
\usepackage{multirow}
\usepackage{array}

\usepackage[colorlinks=true,backref=true,linktocpage=true,citecolor=blue,urlcolor=blue,linkcolor=blue,pdfpagemode=UseOutlines]{hyperref}
\usepackage[dvipsnames]{xcolor}

\begin{document}
\title{
Leveraging neural control variates for enhanced precision in lattice field theory
}

\author{Paulo F. Bedaque}
\email{bedaque@umd.edu}
\affiliation{Department of Physics and Maryland Center for Fundamental Physics, University of Maryland, College Park, MD 20742 USA}

\author{Hyunwoo Oh}
\email{hyunwooh@umd.edu}
\affiliation{Department of Physics and Maryland Center for Fundamental Physics, University of Maryland, College Park, MD 20742 USA}

\begin{abstract}
Results obtained with stochastic methods have an inherent uncertainty due to the finite number of samples that can be achieved in practice. In lattice QCD this problem is particularly salient in some observables like, for instance, observables involving one or more baryons and it is the main problem preventing the calculation of nuclear forces from first principles. 
The method of control variables has been used extensively in statistics and it amounts to computing the expectation value of the difference between the observable of interest and another observable whose average is known to be zero but is correlated with the observable of interest. 
Recently, control variates methods emerged as a promising solution in the context of lattice field theories. 
In our current study, instead of relying on an educated guess to determine the control variate, we utilize a neural network to parametrize this function. Using 1+1 dimensional scalar field theory as a testbed, we demonstrate that this neural network approach yields substantial improvements. Notably, our findings indicate that the neural network ansatz is particularly effective in the strong coupling regime. 
\end{abstract}

\date{\today}

\maketitle

\section{Introduction}

Monte Carlo methods have achieved enormous success in studying non-perturbative field theory phenomena numerically. Still, it faces problems in some models and/or observables where the statistical noise overwhelms the signal. That is the case of theories with a sign problem (see for instance~\cite{Li:2018wnz, deForcrand:2009zkb, Alexandru:2020wrj} for reviews), including all real-time (as opposed to imaginary time) calculations, and models/observables with infinite variance~\cite{Yunus:2022pto, Shi:2015lyu, Alexandru:2022dlq}. It is also the case of certain correlators in lattice QCD whose signal-to-noise ratio decreases exponentially with (imaginary) time, making the extraction of energy levels extremely challenging~\cite{Parisi:1983ae, Lepage:1989hd}. The signal-to-noise ratio of n-baryon states correlators, for instance,  decays as $\sim e^{-( M -3 m_\pi/2)nt}$ ($M$, $m_\pi$ are the baryon and pion masses). Given that excited states contaminate the correlator at small $t$, this decaying signal-to-noise ratio at larger $t$ is a serious problem and is, in fact, the main obstacle one faces in computing the nuclear forces from lattice QCD. 

In this paper we study the use of control variates 
to minimize the variance of lattice observables. The basic idea is simple and well-known~\cite{PhysRevLett.83.4682, Mira_2012}.
The expectation value $\braket{\mathcal{O}}$ of an observable  $\mathcal{O}$ can be computed as $\braket{\mathcal{O}-f}=\braket{\mathcal{O}}$, where $f$ is known to have vanishing expectation value $\braket{f}=0$. While the expectation value of the observables $\mathcal{O}$ and $\mathcal{O}-f$ are the same, their variance is not:
\begin{equation}
    \braket{(\mathcal{O}-f)^2} = \braket{\mathcal{O}^2}+\braket{f^2} - 2 \braket{f\mathcal{O}}.
\end{equation} If a control variate $f$ can be found  that is strongly correlated with $\mathcal{O}$ while maintaining $\braket{f}=0$, the variance of $\mathcal{O}-f$ is smaller than the variance of $\mathcal{O}$ and so it is a better estimator of $\braket{\mathcal{O}}$. This basic strategy has many forms depending on how one goes about finding a suitable $f$. For instance, recently in~\cite{Bhattacharya:2023pxx} control variates for a scalar field theory calculation was found by expressing it as a linear combination of optimal control variates for a free field theory. It has also been argued that control variates can be a possible solution to remove the sign problem exactly~\cite{Lawrence:2020kyw, Lawrence:2022dba}.

In this work, we find control variates for lattice field theory observables using machine learning techniques, namely, a very general function $f$ is parametrized by a feed-forward neural network in such a way that the condition $\braket{f}=0$ is automatically satisfied. Then, standard minimization techniques are used to minimize the variance. In the language of machine learning, the variance becomes the ``cost function'' whose minimization can be thought of as a form of unsupervised learning. The fact that general neural networks are universal function approximators (in the sense of being able to approximate any function with a sufficiently large network) suggests we are looking for a control variate within a very large class of functions. A similar approach has been used in Monte Carlo integration in small dimensional space~\cite{10.1007/978-3-030-46147-8_32, 10.1145/3414685.3417804, sun2023meta, 10.5555/3625834.3625985}.

\section{Method} \label{Method}

The purpose of this section is to review the basic knowledge of the control variates and to explain the way to parametrize the control variates through neural networks. 


\subsection{Control variates} \label{Sec:CV}
Let us denote by $\left\langle \cdot \right\rangle$ the average with respect to the Boltzmann factor:
\begin{equation}
    \left\langle \mathcal{O}(\phi) \right\rangle = \frac{1}{Z}\int D\phi \; \mathcal{O}(\phi) \; {\rm e}^{-S[\phi]},
\end{equation} with $Z=\int D\phi \;  \; {\rm e}^{-S[\phi]}$.
If $\left \langle f \right \rangle = 0$, $\mathcal{O}$ and $\tilde{\mathcal{O}}=\mathcal{O}-f$  will have the same expectation value:
\begin{equation}
    \langle \tilde{\mathcal{O}} \rangle 
   =
   \left \langle \mathcal{O}-f \right \rangle = \left \langle \mathcal{O} \right \rangle. \label{Eq:cvid}
\end{equation}
However, their variances differ:
\begin{equation}
\begin{aligned}
    {\rm Var}(\tilde{\mathcal{O})} & = \left \langle (\mathcal{O}-f)^2 \right \rangle - \left \langle \mathcal{O}-f \right \rangle^2 \\
    & = {\rm Var}(\mathcal{O}) + \left\langle f^2 \right\rangle - 2 \left \langle O f \right \rangle .
\end{aligned}
\end{equation}

The function $f$ is called a control variate. The goal is then to find a function $f$ that is highly correlated with the observable $\mathcal{O}$ in order to minimize the variance of $\tilde{\mathcal{O}}$.

We write our control variate candidate $f$ as
\begin{equation}\label{eq:fdef}
    f(\phi) = \sum_x \left( \frac{\partial g[\phi]_x}{\partial \phi_{x}} -  g[\phi]_x \frac{\partial S}{\partial \phi_{x}} \right) \;,
\end{equation} where $x$ indexes the sites on the spacetime lattice and $g[\phi]:\mathbb{R}^V \rightarrow \mathbb{R}^V$ ($V$ is the spacetime volume) has yet to be defined. For any $g_x[\phi]$, where $g=(g_1, \cdots, g_V)$, integration by parts shows that
\begin{equation}
    \left\langle  \frac{\partial g_x}{\partial \phi_{x}} \right\rangle = \left\langle g_x \frac{\partial S}{\partial \phi_{x}}\right\rangle , \label{Eq:identity}
\end{equation} and therefore,
$\langle f \rangle =0$ by construction.


\subsection{Machine learning} \label{Sec:ML}

While Eq.~(\ref{eq:fdef}) does not define the most general function $f[\phi]$, we aim at having a universal representation of $g[\phi]$. The only constraint we will impose is space-time translational symmetry
\begin{equation}
    f(T_y[\phi]) = f(\phi),
\end{equation} where the translation operator $T_y$ displaces a field configuration by $y$: $T_y[\phi_x] = \phi_{x+y}$.
This can be achieved if $g$ is covariant:
\begin{equation}
    g[T_y[\phi]]_x =  g[\phi]_{x+y}.
\end{equation}

We can impose translational invariance by defining a function $g_0: \mathbb{R}^V \rightarrow \mathbb{R}$ from which we define a $g[\phi]_x$ by 
\begin{equation}
    g[\phi]_x =  g_0(T_{x} [\phi]). \label{Eq:cov}
\end{equation}
It can be easily shown that the control variate $f$ is translational invariant: 
\begin{equation}
    \begin{aligned}
        f(T_y[\phi]) & = \sum_{x} \left(\frac{\partial g[T_y[\phi]]_x}{\partial \phi_{x+y}} -  g[T_y[\phi]]_x \frac{\partial S(T_y[\phi])}{\partial \phi_{x+y}} \right) \\
        & = \sum_{x} \left( \frac{\partial g_0(T_{x+y}[\phi])}{\partial \phi_{x+y}} - g_0(T_{x+y}[\phi]) \frac{\partial S(T_{x+y}[\phi])}{\partial \phi_{x+y}} \right) \\
        & = \sum_{x'} \left( \frac{\partial g_0(T_{x'}[\phi])}{\partial \phi_{x'}} - g_0(T_{x'}[\phi]) \frac{\partial S(T_{x'}[\phi])}{\partial \phi_{x'}} \right) \\
        & = f(\phi) \;.
    \end{aligned} 
\end{equation}

We will define $g_0[\phi]$ by a fully connected feed-forward neural network with $V$ inputs and only one output. For theories with parity symmetry $\phi\rightarrow -\phi$, as the model we will consider in the next section, we choose the activation function $\sigma(x)={\rm arcsinh}(x)$ which is odd.
In addition, we remove the bias term in the linear transformation so that the network is an odd function. In this way, we can reduce the number of parameters, which makes training faster. 

The ideal cost function for the training of $g_0$ is the variance of $\mathcal{O}-f$. We will estimate the variance of $\tilde{\mathcal{O}}$ by the variance of a sample $\{\phi^a \}$, $a=1,\cdots, \mathcal{N}$, of $\mathcal{N}$ field configurations:
\begin{equation}
   L(w) = 
   \frac{1}{\mathcal{N}}\sum_{a=1}^\mathcal{N} \Big(\mathcal{O}(\phi^a) - f_w(\phi^a)\Big)^2,
   \label{Eq:loss1}
\end{equation} where $f_w$, given by Eq.~(\ref{eq:fdef}) and  Eq.~(\ref{Eq:cov}) depends on the parameters $w$ of the network defining $g_0$.
Note that the term from $\langle \mathcal{O} - f \rangle^2$ can be omitted since Eq.~(\ref{Eq:cvid}) does not affect the new variance.

Since, in practice, $\mathcal{N}$ will be small (typically of the order of hundreds in lattice QCD) while the neural network can easily contain a much larger number of parameters, the risk of overfitting exists. In that case, even though $\langle f\rangle=0$ by construction, $\overline{\tilde{\mathcal{O}}} $ is minimized by having $f(\phi^a)\approx\mathcal{O}(\phi^a)$ for every $\phi^a$ in the sample, leading to the sample average $\overline{f} \approx \overline{O}$. This is not a good approximation to the ideal control variate $f$. This problem was recognized in~\cite{10.1007/978-3-030-46147-8_32} where the authors suggest to minimize instead the quantity:
\begin{equation}
   L(w) =
   \frac{1}{\mathcal{N}}\sum_{a=1}^\mathcal{N} \Big(\mathcal{O}(\phi^a) - f_w(\phi^a)-\mu_0\Big)^2.
   \label{Eq:loss2}
\end{equation}
where $\mu_0$ is the sample average of the observable $\mu_0=\overline{O}$. In this case, overfitting leads to $f_w(\phi^a)\approx\mathcal{O}(\phi^a)-\mu_0$ for every $\phi^a$ in the sample and $\overline{f_w} \approx 0$, a much better approximation to the ideal control variate. In fact, \cite{10.1007/978-3-030-46147-8_32, 10.1214/22-BA1328} suggested to use, instead of a fixed value $\mu_0$, a variable $\mu$ that is updated during the training process just like the parameters $w$ of $f_w$. We verified that in our examples this last procedure was significantly better and all results presented below are obtained with this last procedure\footnote{Notice that $\langle \mathcal{O}\rangle$ is still estimated by $\overline{(\mathcal{O}-f_w)}$, {\it not} $\overline{(\mathcal{O}-f_w-\mu)}$.}. Finally, to further avoid overfitting we applied the $L_2$ regularization
\begin{equation}
    L(w, \mu)= \frac{1}{\mathcal N} \sum_{a=1}^{\mathcal N} \left(\mathcal{O}(\phi^a) - f_w(\phi^a) - \mu\right)^2 + \delta \sum w^2, \label{Eq:loss4}
\end{equation}
where $\sum w^2$ is the sum of the squares of all neural network parameters and $\delta$ is the regularization parameter.

\section{Results} \label{Sec:Result}

\begin{figure*}[t]
    \includegraphics[width=1.0\textwidth]{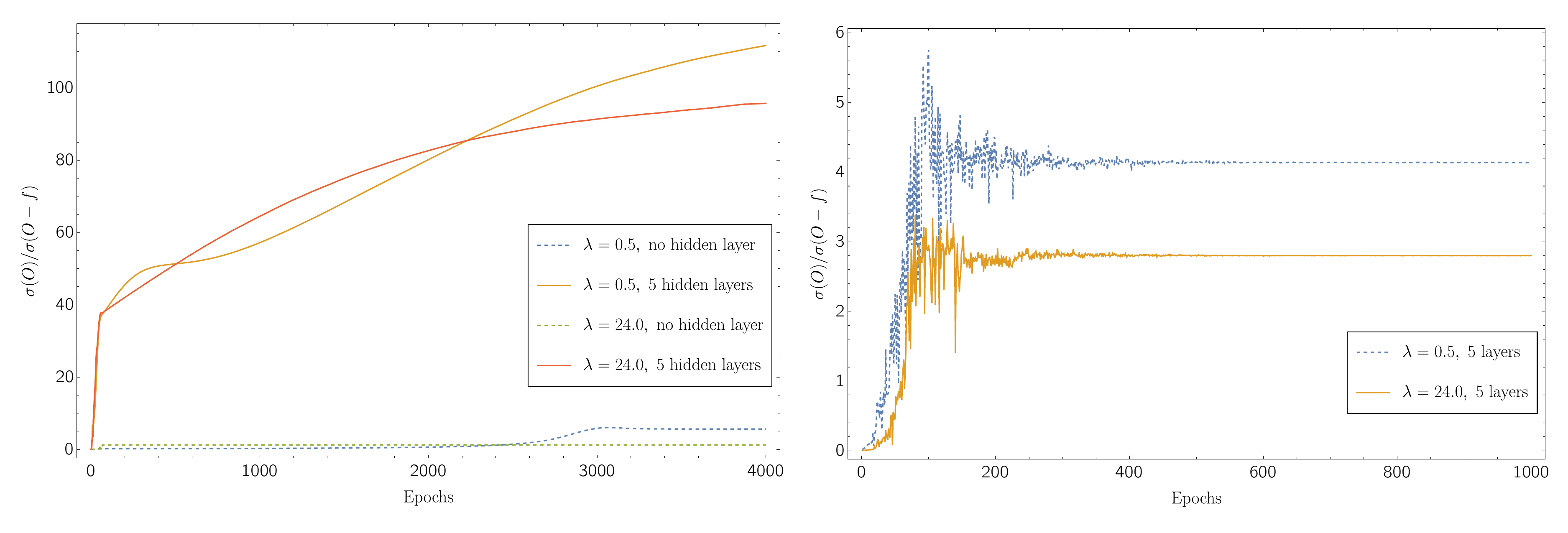}

    \caption{
    Training histories of the small and large couplings on $20 \times 20$ lattice. The figure shows the improvement of standard deviation using control variates with respect to the raw one $(\sigma_{\rm Raw}/\sigma_{\rm CV})$. The dashed line represents the zero hidden layer result (linear transformation), and the solid line represents the result with hidden layers. Networks for small and large couplings have 5 hidden layers and each has 4 neurons. For left and right panels, $10^4$ and $10^3$ samples are reserved for training the network respectively, and $10^3$ samples are used to estimate the variance.
    }
    \label{fig2}
\centering 
\end{figure*}

In order to test our approach 
we use a scalar field theory in 1+1 dimension. Due to its simplicity, the same model has been used as a testbed for other approaches to signal-to-noise problem and sign problem~\cite{Detmold:2018eqd, Detmold:2020ncp, Bhattacharya:2023pxx, Lawrence:2022afv}, so a direct comparison is possible.

We will consider a $L_0 \times L_1$ lattice (So  $V=L_0 L_1$). We work in units where the lattice spacing is  1. Then the action on the lattice can be written as

\begin{equation}
    \begin{aligned}
    S &= \sum_{x, \mu} \Big(  
    \frac{(\phi_{x+\mu}-\phi_{x})^2}{2} 
     + \frac{m^2}{2}\phi^2_{x} + 
     \frac{\lambda}{4!}\phi^4_{x} \Big) \;, 
     \label{Eq:model}
    \end{aligned}
\end{equation} where $\mu=0,1$ indexes the directions on the lattice and $x=(x_0, x_1)$.
The observable we choose to consider is the two-point correlator at momentum $p=0$:
\begin{equation}\label{eq:corr_zeromom}
    \mathcal{O}(t) = \frac{1}{L_0} \sum_{y_0} \left[ \left(  \frac{1}{L_1} \sum_{x_1} \phi_{y_0+t, x_1} \right) \left(  \frac{1}{L_1} \sum_{y_1} \phi_{y_0, y_1} \right) \right].
\end{equation}

\begin{table}[b]
\centering
\begin{tabular}{|| c | c | c | c | c | c | c ||}
    \hline
     & Lattice  & $m^2$   &  $\lambda$ & Layers & Neurons & $\delta$ \\
    \hline
    Fig.~\ref{fig2}  & $20 \times 20$ & 0.1 & 0.5, 24.0 & 5, 5 & 4, 4 & 0 \\
    \hline
    Fig.~\ref{fig4}  & $40 \times 40$ & 0.1 & 0.5, 24.0 & 1, 5 & 1, 32 & 0.05, 0.0005\\
    \hline
    Fig.~\ref{fig:massfit}  & $40 \times 10$ & 0.01 & 0.1 & 1 & 16 & 0.1 \\
    \hline
\end{tabular}
    \caption{Bare parameters of scalar field and neural network parameters in Sec.~\ref{Sec:Result}.} \label{Table:T1}
\end{table}



Table~\ref{Table:T1} summarizes the bare parameters and neural network parameters used in this paper. We use two values of the coupling $\lambda$, couplings $0.5$ and $24.0$ in order to explore  both the weak and strong regimes. The number of hidden layers is chosen from $1$ to $5$, and the number of neurons for hidden layers varies from $1$ to $32$. 
The training of neural networks is performed with the usual stochastic gradient method and the \textit{ADAM} optimizer~\cite{KingBa15} implemented with the help of the  JAX~\cite{jax2018github} and Flax~\cite{flax2020github} libraries.

The choice of learning rate $\eta$ is critical to the success of the training.
If the learning rate is too small, the training might be stuck at local minima. A too large learning rate leads to a process that misses minima of $L(w, \mu)$ altogether. After a lengthy trial-and-error process
we find that it is efficient to use an exponentially decreasing learning rate in the beginning of the training and a constant one after that:
\begin{equation}
    \eta (n) = \begin{cases}
        10^{-3} \times 0.99^{\frac{n}{1000}}, & \text{if $n \le n_0$} \\
        10^{-6}, & \text{if $n \ge n_0$}
    \end{cases} \, , 
\end{equation}
where $n$ is the training step and $n_0$
is in the range $(6.8\times10^5, 6.9\times10^5)$.

Since the gradient of $L(w,\mu)$ is estimated stochastically, we need to decide the number $\texttt{n}$ of samples used. The stochastic error in every step in the gradient descent scales as $\sim \eta/\sqrt{\texttt{n}}$ while the computational cost of following the gradient descent path by unit of ``time" scales as $\sim \texttt{n}/\eta$. Therefore, the optimal choice is $\texttt{n}=1$, the value we used in all our calculations. We have tried other choices, $\texttt{n}\neq1$, and confirmed $\texttt{n}=1$ works better empirically.

\begin{figure*}[t]
    \includegraphics[width=1.0\textwidth]{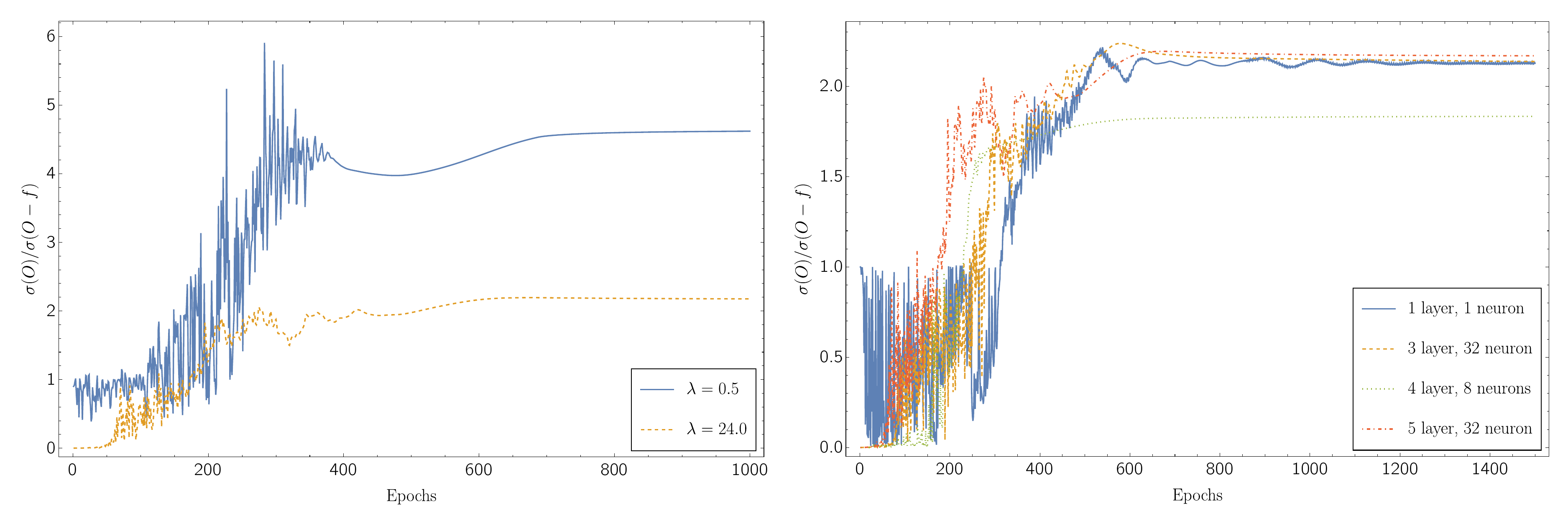}
    
    \caption{Training histories of control variates at $t=L_0/2=20$ on $40\times40$ lattice. The left plot shows the improvement of standard deviation with hidden layers. The right panel displays the training histories of different depths of networks with the large coupling, $\lambda=24.0$. $10^3$ samples are used to train the neural control variates and $10^3$ samples are employed to estimate the standard deviation.
    }
    \label{fig4}
\centering 
\end{figure*}

In Fig.~\ref{fig2} we show the training history on $20 \times 20$ lattice, both at small and large couplings and with networks with and without hidden layers. The observable we aim at improving was a mid-lattice correlator, that is, 
Eq.~(\ref{eq:corr_zeromom}) with $L_0=L_1=20, t=L_0/2=10$.
Fig.~\ref{fig2} shows  the ratio of standard deviations (the improvement in the uncertainty) as a function of the training step for small and large couplings. 
We used $10^4$ and $10^3$ fully decorrelated field configurations to train the network and $10^3$ samples to estimate the variance. An epoch is defined as $\mathcal{N}=10^4$ or $10^3$ gradient descent steps, that is, one step with each training configuration. The result shows that the variance can be greatly improved, but it also implies that the result of training depends on the size of the training set. 
These results can be compared to the ones in~\cite{Bhattacharya:2023pxx} where a different, more direct method was used to obtain control variates for the same theory. Their method is exact for a free theory and it performs better than at weak coupling. 
Neural control variates also work better as weak coupling but outperforms the method in~\cite{Bhattacharya:2023pxx} at strong coupling. As Fig.~\ref{fig2} shows, a linear transformation (network with no hidden layers) performs well at small coupling since a control variate linear on the fields is sufficient for a free theory. 
At strong coupling, the variance is not reduced much with a just a linear combination ansatz, which is shown as the dashed line of figure. However, a better control variate is found by introducing more non-linear terms through increasing hidden layers of neural networks.




\begin{table}[b]
\centering
\begin{tabular}{|| c | c | c | c ||}
    \hline
     & $A$  & $m$   &  $\chi^2/{\rm dof}$ \\
    \hline
    Raw, $\mathcal{N}=2\times10^3$  & 0.000112(18) & 0.1935(40) & 0.55 \\
    \hline 
    CV at $t=20$  &  0.0001191(51) & 0.1920(13) & 0.29 \\
    \hline
    CV at $t=10$  &  0.0001083(44) & 0.1944(13) & 0.55 \\
    \hline
    CV at all points & 0.0001096(7) & 0.1938(2) & 0.91 \\
    \hline
    Raw, $\mathcal{N}=2\times10^5$  & 0.0001088(18) & 0.1940(4) & 0.34 \\
    \hline
\end{tabular}
    \caption{The fitted values of correlators in Fig.~\ref{fig:massfit}.} \label{Table:T2}
\end{table}

The same method, applied to  larger $40\times 40$ lattices, shows a smaller improvement than the $20\times 20$ lattice case, both at weak and strong coupling (see Fig.~\ref{fig4}).  It is notable that networks of different sizes lead to similar results and, in some cases, larger networks do not lead to a larger improvement in the variance. This is in contradiction to the theoretical expectation since, as one increases the number of hidden layers or the number of neurons at each hidden layers, the representability of the neural network increases and, therefore, the larger network should have a better performance than the smaller one. 
Our conclusion is that there is room for improvement in the training of large networks we have not yet able to accomplish. Future work will concentrate on that.

\begin{figure*}[t]
    \includegraphics[width=1.0\textwidth]{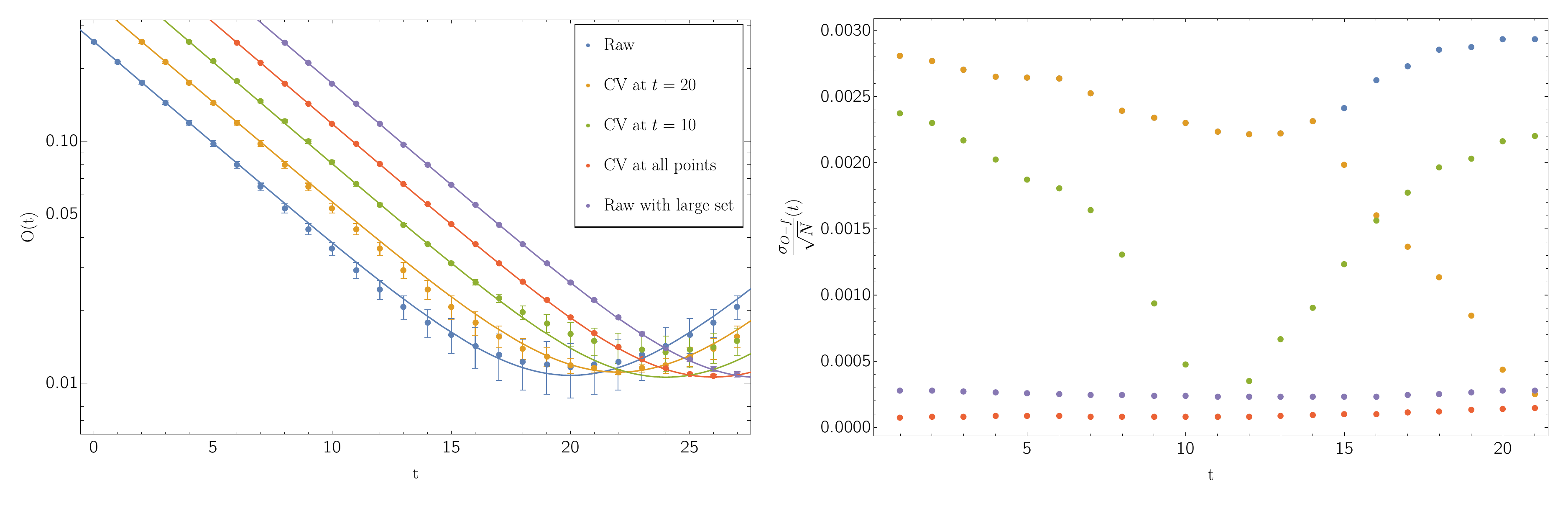}
    
    \caption{Correlation functions with $m^2=0.01$ and $\lambda=0.1$ on $40 \times 10$ lattice. The raw result and the result with control variates are shown. $2\times10^3$ samples are used in total and for the control variate result, $10^3$ samples are used for training and the whole samples are used for estimating observables. For the raw result with large sets, the correlators are calculated with $2 \times 10^5$ samples. The left plot shows the correlation functions with their fitting. Results are shifted horizontally for better visualization. The right plot displays the errors of the correlators in the left plot.
    }
    \label{fig:massfit}
\centering 
\end{figure*}

Reducing the uncertainty of one time slice of a propagator, even by a large factor, doesn't necessarily imply a reduction of the uncertainties of parameters extracted from it, like the value of masses. To investigate this question
we applied our method to anisotropic $40\times 10$ lattices with couplings fairly close to the continuum limit. Ideally, one would use a different control variate at each time slice but this procedure is too expensive. Instead, we found neural control variates optimizing the uncertainty at the  $t=L_0/2=20$ or $t=L_0/4=10$ time slices. The correlators are shown in Fig.~\ref{fig:massfit} and the results of the fits of the correlator to the form
\begin{equation}
    C(t) = A({\rm e}^{- m t} + {\rm e}^{ -m(L_0-t)})
    \label{Eq:massfit}
\end{equation}
are summarized in Table \ref{Table:T2}. Note that the correlation between different points of the correlator is considered when fittings are implemented. 
The regularization strength $\delta$ is chosen as $0.1$. The reduction of the uncertainty due to the use of a control variate is about a factor $\approx 20$ at the time slice $t=L_0/2=20$ (or $t=L_0/4=10$).
Since the data at different time slices are correlated, we find that reducing the variance of one time slice also reduces the variance of nearby time slices, albeit by a smaller amount.
In the example we discussed the uncertainty in the mass estimate is reduced by a factor $\approx 3$ corresponding to a configuration set $\sim 9$ times larger.

While we improved the error around 20 times at one point, the error of the fit giving the mass estimate is  improved by a smaller factor ($\sim 3$). A better mass estimate is obtained if a different control variate is used at each time slice. In order to reduce the computational cost of training we use as a starting point of the training at one time slice, the final result of a previous time slice (transfer training\footnote{We thank the anonymous referee for insisting that we tried this method.}). In doing so, we obtain around 20 times improvement of the standard deviation at every point and the uncertainty of the mass estimate is also improved with the same amount.
We include in Table  \ref{Table:T2} the results of fitting the correlator obtained with much higher statistics in order to verify that the reduced variance result obtained with neural control variates and low statistics is consistent with it.

\section{Discussion} \label{Sec:Discuss}

We showed how control variates parametrized by neural networks reduce the variance in lattice field theories. By using the simple example of a 1+1 dimensional $\phi^4$ theory we established the feasibility of the method and learned some qualitative that can be summarized as: 
\begin{itemize}
    \item Reductions of variance by tens or hundreds are feasible.
    \item It is essential to design the neural network to automatically incorporate the symmetries of the model as those symmetries are difficult to ``learn" from a sample of configurations of a realistic size (hundreds to thousands of configurations). Although we have not explored them yet, other techniques 
    can plausibly work better.
    \item Techniques to avoid overfitting, like the ones we used are essential for good results.
    \item More direct methods, like in~\cite{Bhattacharya:2023pxx}, are more efficient at small coupling but neural networks tend to win out at larger coupling.
\end{itemize}

A natural question is whether how control variates results compare to more standard methods.
After all, the variance of an observable can be reduced by just collecting a larger number $\mathcal{N}$ of configurations. This process is slow as the uncertainties scale as $\sim 1/\sqrt{\mathcal{N}}$. On the other hand, the training of the neural control variate is computationally expensive, even if it needs to be performed only once per observable. The ``break-even" point where our method wins out over the brute force increase in statistics depends on the cost of collecting a new independent configuration. For the scalar theory we studied, the computational cost of the configuration is very small, and for the precision we achieved, it is clearly more efficient simply to increase the statistics. In other theories, specially those with dynamical fermions, the configurations are very expensive and it makes sense to use substantial computing power extracting as much information as possible from them. Thus, the answer to whether (neural) control
variates are more efficient or not has to be studied in every model separately.


Among the direction for future work, the most pressing is the extension of neural control variates to gauge theories. There are conceptual issues to be solved. Given our observation that it is important to impose the symmetries of the model on the neural network, a method to impose gauge invariance seems essential. Also, it is unclear what an analogue to Eq.~(\ref{eq:fdef}) would be. Others issues are of a more practical nature and may be more important. For instance, finding more efficient ways of training neural networks will be crucial to extend the method to realistic, 4-dimensional theories.

\begin{acknowledgments}

This work was supported in part by the U.S. Department of Energy, Office of Nuclear Physics under Award Number(s) DE-SC0021143, and DE-FG02-93ER40762.

\end{acknowledgments}

\bibliography{refs.bib}

\appendix

\end{document}